\documentstyle[12pt]{article}
\input epsf

\begin{document}

\topmargin 0pt
\oddsidemargin 5mm

\setcounter{page}{1}
\begin{titlepage}
\vspace{2cm}
\begin{center}

{\bf Single Charged and Neutral Supersymmetric Higgs Bosons
Production with Jet at pp Colliders}\\
\vspace{5mm}
{\large R.A. Alanakyan,V.H.Grabski}\\
\vspace{5mm}
{\em Yerevan Physics Institute}\\
{Alikhanian Brothers St.2, Yerevan 375036, Armenia\\
alanak @ lx2.yerphi.am}

{
grabsky @ atlas.yerphi.am}
\end{center}

\vspace{5mm}
\centerline{{\bf{Abstract}}}
In the framework of Minimal Supersymmetric Standard
Model Higgs bosons  production via gluino/squark loop in the processes $gu \rightarrow
H^{+}d$, $gd \rightarrow
H^{-}u$,$gq \rightarrow
H^{0}_iq$ are studied. 
\vspace{5mm}
\vfill
\centerline{{\bf{Yerevan Physics Institute}}}
\centerline{{\bf{Yerevan 1997}}}
\end{titlepage}

If Higgs sector of the  gauge theory contains
additonal Higgs fields besides standard doublet,
after spontaneous symmetry violation
  more than one physical Higgs bosons arise.

In particular, in the Minimal Supersymmetric Standard
Model(MSSM)(see \cite{GH}, \cite{HK}   and references therein)
the Higgs sector
contains two doublets of Higgs bosons with opposite hypercharge
(Y=$\pm$ 1) and after spontaneous symmetry violation the
following  physical states appear: charged Higgs bosons
 $H^{ \pm}$, and three neutral ones, $H^0_1, H^0_2, H^0_3 $.

If the  mass of the $H^{\pm}$-boson  is less than $m_t-m_b$, it may
be produced in t-quarks  decays  \cite{GJ}- \cite{DR}:
\begin{equation}
\label{AA}
t\rightarrow H^{+}b
\end{equation}
However, if $m_t-m_b<m_H$ this decay is kinematically forbidden.

That is why it is necessary to study new mechanisms for the
production of heavy
(i.e. with masses $m_t-m_b<m_H$)
charged Higgs bosons.
 
Here we study the following mechanisms  of SUSY  Higgs
bosons production  in $pp$-collisions:

\begin{equation}
\label{AB}
gu \rightarrow H^{+}d,
\end{equation}

\begin{equation}
\label{AC}
gd\rightarrow H^{-}u ,
\end{equation}

\begin{equation}
\label{AD}
gq \rightarrow H^{0}_i q
\end{equation}

which proceed via squark/gluino loop (Fig.1) .

It must be noted that besides the  squark/gluino loop
contribution there is also the tree contribution \cite{BHS}. It is
of interest to compare both contributions.The cross section of
the tree contribution to the subprocess (2) is of order
$O(\alpha_s \alpha \frac{m_q^2}{m_W^2} \tan^2 \beta)$, whereas the
 loop contribution is of order $O(\alpha_s^3 \alpha
  \sin^22 \beta)$.Besides the tree contribution is supressed
  by the smallness of the heavy quarks (s,c,b,...)  inside
  protons.Thus we expect that the loop contribution will dominate over
 the tree at not
 very large $\tan \beta$ .
It must be noted that besides the squark/gluino loop
contribution there are also contributions from squarks and
t-quarks loops in the process (4).The heavy quark and squark
contribution in the process (4) and the process
:
\begin{equation}
\label{AE}
gg\rightarrow H^{0}_i g
\end{equation}
has been considered in \cite{CLLR}- \cite{DK}.

Using Higgs bosons interactions with scalar quarks (see formula
(4.19)
in ref. \cite{GH}) :
\begin{eqnarray}
\label{AF}
&& L=- \frac{g}{\sqrt{2}}m_W \sin 2 \beta(H^+\tilde {u_L^*}
\tilde{d_L}+H.c.)-\frac{gm_Z}{\cos \theta_W} \sum((T_{3i}-e_i
\sin^2\theta_W)^2 \tilde{q_{iL}^*}
\tilde{q_{iL}}+\nonumber\\
&& e_i \sin^2 \theta_W)
\tilde{q_{iR}^*}
\tilde{q_{iR}}) (H^{0}_1 \cos(\alpha+ \beta)-H^{0}_2
\sin(\alpha+ \beta))
\end{eqnarray}
for the gauge invariant amplitude of the process (2) we obtain:
\begin{equation}
\label{AW}
M= \sqrt{2} \frac{\alpha_s^{\frac{3}{2}}\alpha^{\frac{1}{2}}}{\sin\theta_W}m_W
 \sin 2 \beta \bar{u}(k_1)T^a \gamma_{\mu}
P_Lu(k_2)(G_{\mu \nu}^a(F_1k_1^{\nu}+F_2k_2^{\nu})+
\epsilon_{\mu \nu \lambda \rho}G_{\lambda \rho}^a(F_3k_1^{\nu}+F_4k_2^{\nu})).
\end{equation}

Here

\begin{equation}
\label{AG}
G_{\mu \nu}^a=k_{\mu}A^a_{\nu}-k_{\nu}A^a_{\mu}
\end{equation}

\begin{equation}
\label{AH}
F_1=Nf_3+\frac{1}{N}f_1,
\end{equation}

\begin{equation}
\label{AI}
F_2=Nf_4+\frac{1}{N}f_2 ,
\end{equation}

\begin{equation}
\label{AJ}
F_3=\frac{N}{2}\int\limits_{0}^{1}dx\int\limits_{0}^{1-x}dy A_3 ,
\end{equation}

\begin{equation}
\label{AK}
F_4=\frac{N}{2}\int\limits_{0}^{1}dx\int\limits_{0}^{1-x}dy xD ,
\end{equation}
(where $N=3$ number of colours),

\begin{equation}
\label{AL}
f_1=\int\limits_{0}^{1}dx\int\limits_{0}^{1-x}dy(-(x+y)A_1-yA_2),
\end{equation}

\begin{equation}
\label{AM}
f_2=\int\limits_{0}^{1}dx\int\limits_{0}^{1-x}dy((x+y)A_2-yA_1),
\end{equation}

\begin{equation}
\label{AN}
f_3=\int\limits_{0}^{1}dx\int\limits_{0}^{1-x}dy(x+y)(-A_3),
\end{equation}

\begin{equation}
\label{AO}
f_4=\int\limits_{0}^{1}dx\int\limits_{0}^{1-x}dyx(1-x-y)D,
\end{equation}

\begin{equation}
\label{AP}
A_i= \frac{1}{a_i^2}
(\log(\frac{a_i(1-x-y)+b_i}{b_i})+\frac{a_i(1-x-y)}{a_i(1-x-y)+b_i}),
\end{equation}

\begin{equation}
\label{AQ}
D=\frac{1-x-y}{b_3(a_3(1-x-y)+b_3)},
\end{equation}

\begin{equation}
\label{AR}
a_1=(m_H^2-t)y+sx,
\end{equation}

\begin{equation}
\label{AS}
b_1=-m^2_{\tilde{g}}x-m^2_{\tilde{u}}(1-x)+ty(1-x-y)+i\epsilon,
\end{equation}

\begin{equation}
\label{AT}
a_2=(m_H^2-t)y+ux,
\end{equation}

\begin{equation}
\label{A1}
b_2=b_1,
\end{equation}
\begin{equation}
\label{A2}
a_3=(m_H^2-u)x+sy+m^2_{\tilde{g}}-m^2_{\tilde{u}},
\end{equation}

\begin{equation}
\label{A3}
b_3=-m^2_{\tilde{g}}(1-x)-m^2_{\tilde{u}}x+ux(1-x-y)+i\epsilon.
\end{equation}

In our calculations we use $m_{\tilde{u}}=m_{\tilde{d}}$ approximation
 because in MSSM for left scalar
quarks (if we neglect masses of the light quarks) we have:

\begin{equation}
\label{A4}
m^2_{ \tilde{u}}-m^2_{ \tilde{d}}=( 1- \sin^2 \theta_W)m^2_Z\cos{2\beta}.
\end{equation}

The processes of the neutral Higgs bosons production via
gluino/squark loop may be obtained from formulas (7)-(21), taking into
account (6), by the following replacements:

\begin{equation}
\label{A5}
m_W\sin(2\beta)\rightarrow m_Z((T_{3i}-e_i
\sin(\theta_W))^2)\tilde{q_{iL}}
\tilde{q_{iL}}+e_i
\sin(\theta_W))^2)\tilde{q_{iR}} \tilde{q_{iR}} \cos(\alpha+ \beta)
\end{equation}

\begin{equation}
\label{A6}
m_W\sin(2\beta)\rightarrow m_Z((T_{3i}-e_i
\sin(\theta_W))^2)\tilde{q_{iL}}
\tilde{q_{iL}}+e_i
\sin(\theta_W))^2)\tilde{q_{iR}} \tilde{q_{iR}} \sin(\alpha+ \beta)
\end{equation}

for $H_{1,2}^0$ -bosons respectively.In the case of the charged
Higgs bosons production only left scalar quarks in loop
contribute to the processes (2),(3).It must be noted also that in
case of the neutral scalar Higgs bosons production in the
subprocess (4) the total amplitude is the sum of both pure
both left and right scalar quarks loop contributions, the
pure$t$-quark loop contribution and the gluino/squark loop
contribution.The tree and loop contribution do not interfere with
each other.

It must be noted also ,in case of the neutral Higgs bosons
production the total amplitude is the sume of the gluino/squark
amplitude and pure t-quark and squark amplitude.

For the differential cross section  of the subprocess (2) we obtain
the following result:

\begin{equation}
\label{A7}
\frac{d\sigma }{dt}=-\frac{\alpha \alpha^3_s m_W^2}{128
\pi \sin^2\theta_W} (\sin 2\beta)^2 \frac{t}{s}(s^2 (\mid (F_1
\mid ^2+\mid F_3 \mid ^2)+
u^2 (\mid F_2 \mid^2+\mid F_4\mid ^2))
\end{equation}

Here we  use the following notations:
 $s=(k_1+k)^2$,
 $t=(k_1-k_2)^2$,
 $u=(k_1-k_3)^2$,
 $m_H$ is the  mass of $H^{+}$ or $H^{0}_i$-bosons,

\begin{equation}
\label{A8}
\sigma(pp \rightarrow H^+ +jet+X)=
\int\limits_{\frac{m_H^2}{s_0} }^{1}dx_1
\int\limits_{\frac{m_H^2}{x_1s_0}}^{1}dx_2
\sigma(x_1x_2s_0)(u(x_1)g(x_2)+u(x_2)g(x_1))
\end{equation}

with replacements

\begin{equation}
\label{A9}
u(x) \rightarrow d(x),u(x)+d(x)
\end{equation}

for $H^-$ and $H^0_i$ Higgs bosons respectively.
For structure functions we use parametrizations of ref. \cite{E}.

 The numerical result will be presented in the nearest
 future. The work in this direction is in progress.

The authors express their sincere gratitude to H.Hakopian for
helpful discussions.

\newpage

\begin{figure}
\begin{center}
\epsfxsize=10.cm
\leavevmode\epsfbox{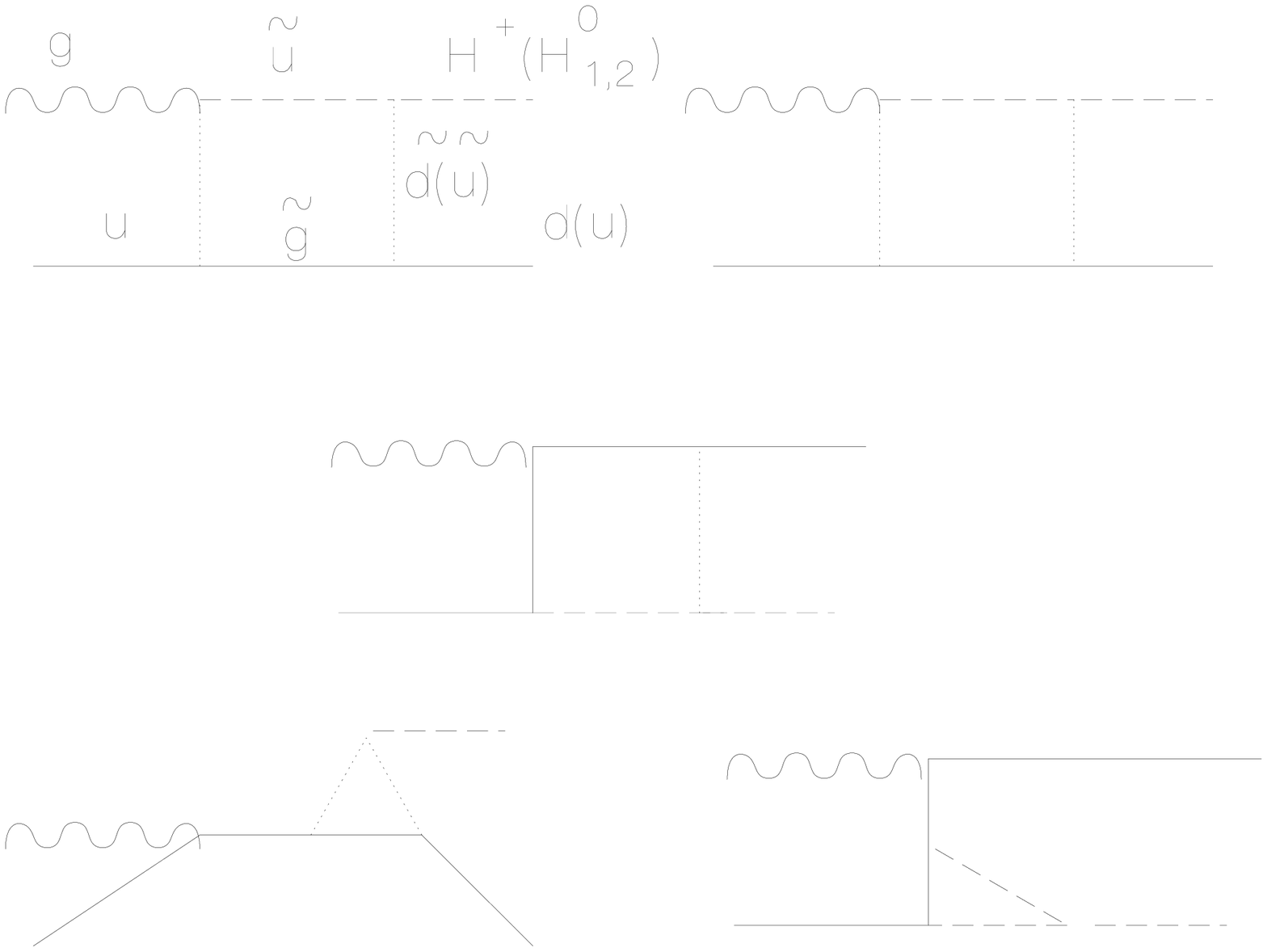}
\end{center}
\caption{
Diagrams corresponding to the processes (2)-(4).
}
\end{figure}

\end{document}